# Understanding the Results of Electrostatics Calculations: Visualizing Molecular 'Isopotential' Surfaces

Cameron Mura[1]*

[1]Department of Chemistry; University of Virginia; Charlottesville, VA 22904 USA; *cmura@virginia.edu

July 2014

## Preliminary notes & motivation

This document attempts to clarify potential confusion regarding electrostatics calculations, specifically in the context of biomolecular structure. We begin with some notes:

- The results of a **molecular electrostatics calculation** can be visualized in many ways — as isopotential surfaces or equipotential contours, as electrostatic potentials mapped onto a solute's van der Waals surface, solvent-accessible surface, *etc.* Such illustrations are becoming increasingly common in the biochemical literature; some examples are shown below.

- To learn more about actually performing the electrostatics calculation and loading the raw results (*e.g.*, an OpenDX-formatted volumetric map) into, say, a PyMOL session (using PyMOL's APBS plugin), one can visit http://dx.doi.org/10.1371/journal.pcbi.1000918 and click on "Video S2" – Electrostatics screencast (near the bottom; see ref [4] for more on this). Also, it would be a good idea to visit the APBS site, http://www.poissonboltzmann.org/apbs, and familiarize oneself with the underlying principles and practical usage of this software.

- An electrostatics calculation package, sometimes termed an electrostatics '*solver*' (*e.g.* APBS [1], DelPhi), computes an approximate numerical solution to the Poisson-Boltzmann equation given a specific, well-defined *problem description* — *i.e.*, *(i)* a particular three-dimensional (3D) molecular geometry, consisting of atomic coordinates for solute atoms, ions, and whatever other species will be treated explicitly in the calculation (*e.g.*, a small-molecule ligand), and *(ii)* bulk properties for the system (ionic species and concentrations, solute and solvent dielectric constants, and so on). Such calculations result in electrostatic potentials, which are inherently a form of *volumetric* data — that is, the output is a set of numerical values of some dependent variable (our function), $f$, which varies with position in space, $\vec{r}$, that is $f(\vec{r})$. The function $f$ depends upon one or more *independent variables* by which we mean, for instance, position in 3-space, $\vec{r} \in \Re^3$ (or whatever region we take as our problem domain).

  This function $f$ may itself be *vector-valued*, such as electric field lines (or really any *force field*). Or, the function may be *scalar-valued*, such as when the function under consideration is the electrostatic potential. For more background, recall from physics and vector calculus [5] that many important vector fields that arise in nature — the *gravitational force field*, the *electric field* (or *electrostatic field* if we consider only *stationary* charges [time-independent]), *etc.* — are described as *conservative* vector fields (or *conservative* functions/forces because they conserve energy, are *path-independent*, *etc.*)... and, in such cases these vector fields are expressible as the gradients of some type of function (think of it as the *antiderivative*), which is known in this context as a *scalar potential*, *P*. In a nutshell, the relationship is that the vector field function, $\vec{F}$, is the negative gradient of the potential: $\vec{F} = -\nabla P$. So, for example, *(i)* the electric(static) field is the negative gradient of the electric(static) potential ($\vec{E} = -\nabla \Phi$, for electrical potential $\Phi(\vec{r}) = U_{el}(\vec{r})/q$)), *(ii)* the gravitational field experienced at some position $\vec{r}$ is the negative gradient of the gravitational potential at that position, and so on.

[Note that the electron density, $\rho(\vec{r})$, familiar to crystallographers, is an example of scalar-valued volumetric data.]

- *What do images such as the ones below really* **mean**? – Answering this question is the basic purpose of this document.

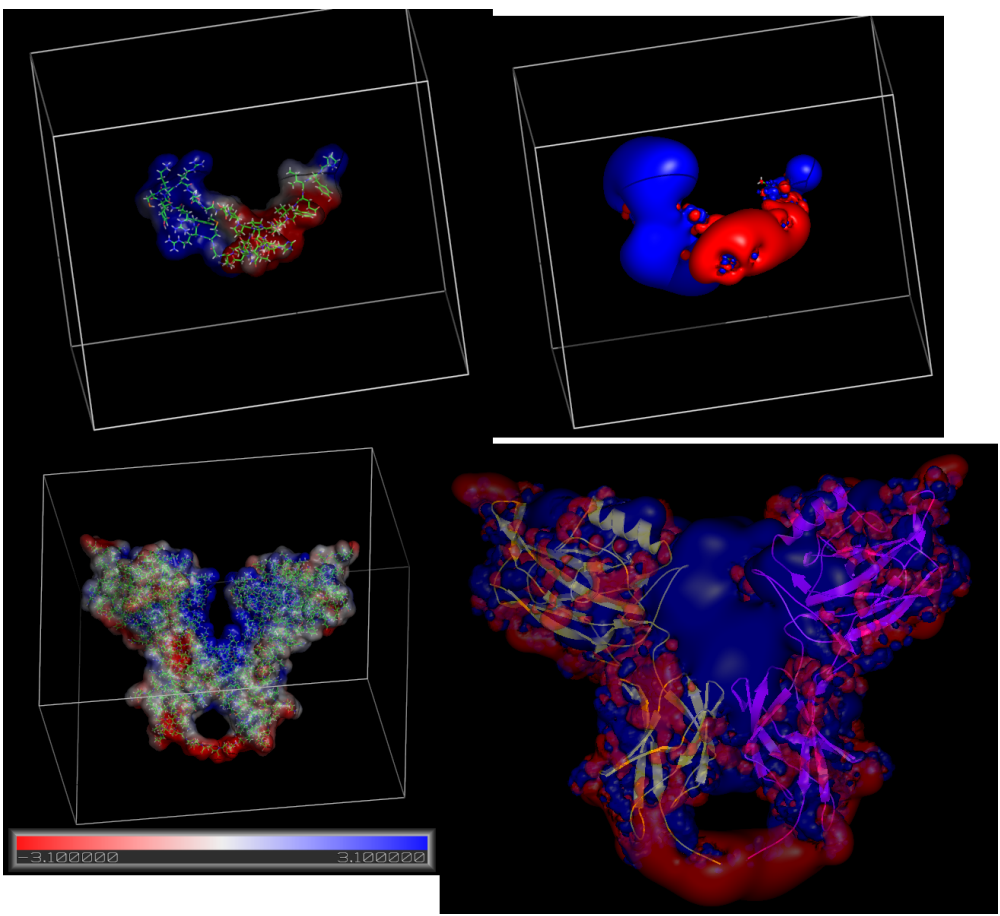

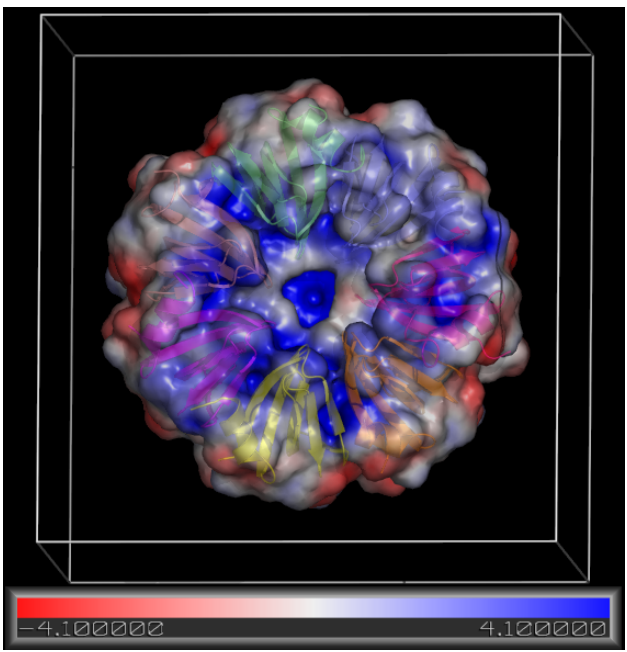

## What do these illustrations mean?

Specifically, what is the significance/meaning of the magnitudes (and units) of the red/blue scalebars that are often shown alongside these images, such as some of the above illustrations? What are 'appropriate' min/max values for displaying the potential?

**Regarding cutoffs**: For visualizing the output of electrostatics calculations, appropriate min/max threshold values are generally $\pm 5 k_\text{B}T/e$, or thereabouts. The 'thereabouts' is intended not to be vague, but rather because there is no universal 'rule' for this: Of course, contouring between $\pm 10$ $k_\text{B}T/e$ is more stringent than $\pm 5$ (the surfaces will be more intense in the latter [maybe misleadingly so?]), so if that is what one would like to convey with an image that is fine too, so long as these parameters are properly noted in the accompanying caption. The main trickery to beware of is if someone, say, *(i)* shows values at $\pm 1\ k_\text{B}T/e$, *(ii)* compares this electrostatics output to experiments in high-[salt] buffer, and then *(iii)* claims something like "Hey look at that strong positively charged surface patch, it must interact with RNA!" [3].

**Regarding units**: At http://muralab.org/~cmura/PyMOL/Electrostatics/apbs_plugin.html, there occur statements such as "Same potentials as above, displayed as isopotential surfaces contoured at levels of $+2k_\text{B}T/e$ (blue) and $-2k_\text{B}T/e$ (red)". A question often asked about these continuum electrostatics calculations is "What does this '$k_\text{B}T/e$' terminology mean?" What it means is that the electrostatic component of the potential energy of interaction between *(i)* the field due to the biomolecule and *(ii)* an elementary charge of $+1e$ (*i.e.*, a proton), which is located somewhere on the (say) $+2k_\text{B}T/e$ isopotential surface, would equal:

    == (2 *   $k_\text{B}T$)   / (+1 * 1.602E-19 Coulombs)
    == (2 * (8.314 Joules/Kelvin / 6.022E+23) * 298 K) / (+1 * 1.602E-19 Coulombs)
    == +0.0514 *Volts* (recall 1 *V* = 1 J/C)

                3 of 4

NB: In the above explanation, the '$k_\text{B}T/e$' is not a unitless quantity, contrary to occasional misconceptions. This quotient would be unitless if you were to view the '$e$' as being in an energy unit, such as electron volts ($eV$)...then, '$k_\text{B}T$' (or '$RT$' for the molar quantity) works out to be 2.48 kJ (per mole) at room temperature, and 1 $eV = 1.602*10^{19}$ J (per particle). So, in that way you could (wrongly) view the units as 'cancelling', but this really is **not** the meaning of $k_\text{B}T/e$ — rather, it is energy (in fact the fundamental scale for thermal energy, '$k_\text{B}T$'), normalized by charge ('$e$'), and that quotient (energy per unit charge) is the definition of *voltage*. Also, as another way of thinking about it, the above development is akin to the relationship between an electrostatic force at some point in space ($\vec{F}_{el}$) and the magnitude of the electric field at that point ($\vec{E}$): $\vec{E} = \vec{F}_{el}/q$ for charge '$q$'. In other words, one can think of the electrostatic field [2] as being the electrostatic force normalized by an elementary unit of test-charge (...and so $k_\text{B}T/e$ can be viewed as an *energy density*).

## Acknowledgements

This work in the Mura lab at UVa was partly supported by UVa start-up funds, the Jeffress Memorial Trust (J-971), and an NSF Career award (MCB-1350957).